\documentclass[referee,structabstract]{aa}
\usepackage{graphicx}
\usepackage{txfonts}
\usepackage{hyperref}
\hypersetup{
    colorlinks,
    citecolor=blue,
    filecolor=blue,
    linkcolor=blue,
    urlcolor=blue,
    menucolor=black}
\begin{document}

\newcommand{\nmag}{n_{\rm mag}}
\newcommand{\rdnmag}{\Delta\nmag}
\newcommand{\srdnmag}{\left<\rdnmag\right>_{\rm 181d}}

\newcommand{\blos}{B_{\rm los}}
\newcommand{\rdblos}{\Delta\blos}
\newcommand{\srdblos}{\left<\rdblos\right>_{\rm 181d}}
\newcommand{\rmsblost}{\sigma_{\blos{\rm,time}}}
\newcommand{\rblos}{R^2_{\blos}}
\newcommand{\mblos}{m_{\blos}}
\newcommand{\rmsblos}{\sigma_{\blos}}
\newcommand{\rmsblosa}{\sigma_{\blos{\rm,A}}}
\newcommand{\rmsblosb}{\sigma_{\blos{\rm,B}}}
\newcommand{\rmsblosabs}{\sigma_{\blos{\rm,abs}}}
\newcommand{\rmsblosmm}{\sigma_{\blos{\rm,mag,ddc}}}

\newcommand{\ftot}{\phi_{\rm tot}}
\newcommand{\ftotfn}{\phi_{\rm tot,fn}}
\newcommand{\rdftot}{\Delta\ftot}
\newcommand{\srdftot}{\left<\rdftot\right>_{\rm 181d}}
\newcommand{\rmsftott}{\sigma_{\ftot{\rm,time}}}
\newcommand{\rftot}{R^2_{\ftot}}
\newcommand{\mftot}{m_{\ftot}}
\newcommand{\rmsftot}{\sigma_{\ftot}}
\newcommand{\rmsftota}{\sigma_{\ftot{\rm,A}}}
\newcommand{\rmsftotb}{\sigma_{\ftot{\rm,B}}}
\newcommand{\rmsftotmm}{\sigma_{\ftot{\rm,mag,ddc}}}

\newcommand{\ablos}{\left|B_{\rm los}\right|}

\newcommand{\inmf}{\left<\ablos/\mu\right>_{\rm in}}

\title{The variation in the response of solar full-disc magnetographs}
\titlerunning{The variation in the response of solar full-disc magnetographs}
\author{K. L. Yeo\inst{1} \and S.~K.~Solanki\inst{1,2} \and N.~A.~Krivova\inst{1}}
\institute{Max-Planck Institut f\"{u}r Sonnensystemforschung, Justus-von-Liebig-Weg 3, 37077 G\"{o}ttingen, Germany\\ \email{yeo@mps.mpg.de} \and School of Space Research, Kyung Hee University, Yongin, 446-701 Gyeonggi, Korea}

\abstract{The utility of full solar disc magnetograms as a long-term record of the photospheric magnetic field requires an understanding of how stable these observations are with time and the systematic differences between the various instruments.}{We compared magnetograms from the KPVT/SPM, SoHO/MDI, SOLIS/VSM, and SDO/HMI with the aim of probing the effect on measured solar magnetism of the variation in instrument response with time, magnetogram signal level, and position on the solar disc.}{Taking near-simultaneous observations from the various instruments, we examined the surface coverage by magnetic activity and the effect of cross-calibrating the various instruments under different assumptions.}{By comparing the surface coverage by magnetic activity in the observations from the various instruments, we traced the effect of the time variation in instrument response on the longitudinal magnetogram signal and disc-integrated unsigned magnetic flux. This yielded evidence of acute changes in the response of MDI and VSM with certain events such as the SoHO vacation in 1998 and the upgrade of the VSM CCD camera in 2009. Excluding these changes, the effect of instrument instability on the magnetogram signal and disc-integrated magnetic flux appears to be rather benign, with an associated uncertainty of less than 2$\%$. We determined the magnetogram signal ratio between each instrument pairing as a function of magnetogram signal level and distance from disc centre and with it cross-calibrated the various instruments. We compared the result with that from repeating the cross-calibration with the overall magnetogram signal ratio. This allowed us to estimate the uncertainty in the magnetogram signal associated with the variation in instrument response with magnetogram signal level and distance from disc centre to be about 8$\%$ to 14$\%$. The corresponding uncertainty in the disc-integrated magnetic flux is about 7$\%$ to 23 $\%$.}{To the best of our knowledge, this is the first study of its kind to quantify the uncertainty in measured magnetism from the variation in instrument response with time, magnetogram signal level, and disc position. The results here will be useful to the interpretation of SPM, MDI, VSM, and HMI magnetograms. As examples, we applied our findings to selected results from earlier studies based on such data.}
\maketitle

\section{Introduction}
\label{introduction}

Since the 1960s, a succession of solar telescopes recorded longitudinal magnetograms of the full solar disc at sub-minute to daily intervals. This body of observations, giving the spatial distribution and time evolution of the photospheric magnetic field over the past half-century, is of obvious importance to the understanding of solar magnetism. Through its application to models of space weather \citep{toth05,bobra15} and solar irradiance variability \citep{krivova03,wenzler04,wenzler06,ball12,yeo14}, which provides the solar forcing input required by climate simulations \citep{haigh07,ermolli13,solanki13}, this body of observations is also pertinent to the study of Sun-Earth interactions.

Measured photospheric magnetic flux density, i.e. the magnetogram signal, is modulated by instrumental factors such as the design, calibration, stray light, spectral line surveyed, spatial and spectral resolution, and specifics of the data reduction. The result is the systematic differences between the observations from the various instruments. This is most readily apparent in the conversion factor, the ratio of the magnetogram signal registered by two given instruments, deviating from unity \citep[see Fig. 9 in][]{riley14}. Reconciling the observations from the various instruments is complicated by the time variation in instrument response, which can arise from hardware changes, thermal fluctuations, and ageing and exposure degradation.

Few studies have examined the stability of full-disc magnetographs. \cite{arge02} compared synoptic magnetograms from the Mount Wilson Observatory, Wilcox Solar Observatory, and Kitt Peak Vacuum Telescope (KPVT). The authors looked at the total photospheric magnetic flux and open magnetic flux and noted that while the various data sets are broadly consistent with one another, there are certain discrepancies which they attributed to instrument changes. \cite{riley14} extended this study by including similar observations from the Global Oscillation Network Group \citep[GONG,][]{harvey96}, the Michelson Doppler Imager onboard the Solar and Heliospheric Observatory \citep[SoHO/MDI,][]{scherrer95} and the Helioseismic and Magnetic Imager onboard the Solar Dynamics Observatory \citep[SDO/HMI,][]{scherrer12}. The authors examined the conversion factor as a function of time to trace the relative instrument response. Similar to \cite{arge02}, \cite{riley14} reported that while the conversion factor between the various magnetographs appears to be broadly stable with time, there are indications of instrument changes. While such studies have highlighted instrument changes, the effect on measured magnetism, such as the observed magnetic flux density and amount of magnetic flux, has not been investigated in detail.

Besides \cite{riley14}, a number of other studies examined the conversion factor between the various full-disc magnetographs. The conversion factor was found to vary with magnetogram signal level \citep{jones01,wenzler04,liu12,pietarila13,riley14,yeo14} and position on the solar disc \citep{tran05,demidov08,pietarila13,riley14,yeo14}. Ideally, the magnetogram signal at a given point should scale linearly with the actual magnetic flux density there and be independent of the location of the point on the solar disc. The variation in the conversion factor with magnetic flux density and disc position suggest however, that instrument response fluctuates such that the magnetogram signal at a given point can scale nonlinearly with the actual magnetic flux density there and change with the location of the point on the solar disc. This can arise from inhomogeneities in instrumental properties across the image plane and uncertainties in the calibration and data reduction. As with the time variation in instrument response, the effect of the changes in instrument response with magnetogram signal level and disc position on measured magnetism has not been investigated in detail.

The variation in instrument response with time, magnetogram signal level and disc position is relevant to the reconciliation of the observations from the various full-disc magnetographs and in consequence the utility of this body of observations as a long-term record of photospheric magnetism. In this study, we aim to address the issue that the effect of such fluctuations in instrument response on measured magnetism is not well studied. For this purpose, we compare longitudinal magnetograms from the higher spatial resolution (pixel scale of 2 arcsec and finer) full-disc magnetographs that had operated over the past three decades (Table \ref{instruments}): the KPVT SPectroMagnetograph \citep[SPM,][]{jones92}, SoHO/MDI, the Vector SpectroMagnetograph in the Synoptic Long-term Investigations of the Sun instrument suite \citep[SOLIS/VSM,][]{keller03}, and SDO/HMI. From an examination of the solar disc coverage by magnetic features and the conversion factor between the various instruments, we will estimate the uncertainty in the observed magnetic flux density and disc-integrated photospheric magnetic flux associated with the fluctuations in instrument response. In the following, we will describe the data selection and reduction (Sect. \ref{data}) and the aforementioned analysis (Sect. \ref{analysis}), before a discussion of the results  (Sect. \ref{discussion}) and a summary of the study (Sect. \ref{summary}).

\begin{table*}
\caption{Overview of the instruments examined in this study.}
\label{instruments}
\centering
\begin{tabular}{ccccc}
\hline\hline
& & & Image size & Pixel scale\tablefootmark{a} \\
Instrument & Period [year.month.day] & Spectral line & [pixel] & [arcsec] \\
\hline
SPM                  & 1992.11.19 - 2003.09.21 & Fe I 8688 \AA{} & $1788\times1788$ & 1.14  \\
MDI                  & 1996.03.19 - 2011.04.11 & Ni I 6768 \AA{} & $1024\times1024$ & 1.98  \\
VSM\tablefootmark{b} & 2003.08.23 - 2009.11.27 & Fe I 6302 \AA{} & $2048\times2048$ & 1.125 \\
                     & 2009.12.18 - 2017.10.27 &                 &                  & 1     \\
HMI                  & 2010.04.30 - present    & Fe I 6173 \AA{} & $4096\times4096$ & 0.504 \\
\hline
\end{tabular}
\tablefoot{\tablefoottext{a}{The nominal spatial resolution is, for each instrument, about twice the pixel scale, though for ground-based telescopes such as SPM and VSM, this spatial resolution is never reached due to atmospheric seeing.} \tablefoottext{b}{The CCD camera of VSM was upgraded in the gap between 2009.11.27 and 2009.12.18, giving the change in pixel scale.}}
\end{table*}

\section{Data selection and reduction}
\label{data}

In this study, we compare full-disc longitudinal magnetograms from KPVT/SPM \citep{jones92}, SoHO/MDI \citep[fd\_M\_96m\_lev182 data series,][]{scherrer95}, SOLIS/VSM \citep{keller03}, and SDO/HMI \citep[M\_720s data series,][]{schou12}. In Table \ref{instruments}, we list the period of operation, spectral line surveyed, image size, and pixel scale of each instrument.

\subsection{Exceptional instrument events}
\label{instrumentevents}

First, let us describe certain events that occurred to MDI and VSM, which are exceptional in that they are outside the normal operation of such instruments, that altered their spatial resolution (c.f., Sects. \ref{preparation} and \ref{analysisstability}). We exclude these events from the assessment of instrument stability by considering MDI and VSM observations from before and after each event separately.

The SoHO spacecraft experienced a series of extended outages between June 1998 and February 1999, often referred to as the SoHO vacation, during which MDI might have degraded \citep[see discussion in][]{ball12}. Also, the instrument was refocused after the SoHO vacation. Let us denote MDI before and after the SoHO vacation and the refocus of this instrument as MDI(1) and MDI(2), respectively.

The CCD camera of VSM was upgraded in December 2009, changing the pixel scale from 1.125 arcsec to 1 arcsec (Table \ref{instruments}). The instrument team scaled the pre-upgrade magnetograms by a factor of 1.59 to reconcile them to the post-upgrade magnetograms. However, \cite{pietarila13} demonstrated that even with this step, systematic differences remain between the pre-upgrade and post-upgrade magnetograms. In another change, the VSM was operated with the interior filled with helium till January 2014, after which nitrogen was employed instead. This resulted in a weak but palpable degradation in spatial resolution \citep{harvey14}. Here, we will refer to VSM before the CCD camera upgrade as VSM(1), from between this and the switch to nitrogen as VSM(2), and from after the switch as VSM(3).

\subsection{Data selection}
\label{selection}

Considering MDI and VSM observations from before and after the exceptional instrument events we had just discussed (Sect. \ref{instrumentevents}) separately, we make the following instrument comparisons: MDI(1)-SPM, MDI(2)-SPM, MDI(2)-VSM(1), MDI(2)-VSM(2), HMI-VSM(2), HMI-VSM(3), and HMI-MDI(2) (Table \ref{instrumentpairs}). For each instrument comparison, for each day on which both instruments recorded observations, we select the magnetogram from each instrument taken closest in time to one another. The result is a set of daily near-simultaneous magnetogram pairs.

\begin{table*}
\caption{Overview of the instrument comparisons.}
\label{instrumentpairs}
\centering
\begin{tabular}{ccccccc}
\hline\hline
& Number of daily & & & FWHM\tablefootmark{b} & \multicolumn{2}{c}{Magnetogram} \\
Comparison & magnetogram pairs & Period [year.month.day] & Coverage\tablefootmark{a} & [arcsec] & \multicolumn{2}{c}{noise threshold\tablefootmark{c} [G]} \\
\hline
MDI(1)-SPM    & 87  & 1996.04.23 - 1998.06.08 & 0.112 & 4.77 & MDI(1): 16.6 & SPM: 13.4    \\
MDI(2)-SPM    & 182 & 1999.02.25 - 2003.09.05 & 0.110 & 4.44 & MDI(2): 23.1 & SPM: 15.8    \\
MDI(2)-VSM(1) & 385 & 2003.09.01 - 2009.11.20 & 0.169 & 5.44 & MDI(2): 18.8 & VSM(1): 19.3 \\
MDI(2)-VSM(2) & 104 & 2009.12.18 - 2011.04.11 & 0.217 & 6.34 & MDI(2): 14.9 & VSM(2): 13.7 \\
HMI-VSM(2)    & 406 & 2010.05.01 - 2014.01.28 & 0.297 & 5.41 & HMI: 9.09    & VSM(2): 11.6 \\
HMI-VSM(3)    & 490 & 2014.02.04 - 2017.10.20 & 0.362 & 6.04 & HMI: 10.6    & VSM(3): 14.0 \\
HMI-MDI(2)    & 287 & 2010.04.30 - 2011.04.11 & 0.827 & 3.32 & HMI: 17.6    & MDI(2): 27.3 \\
\hline
\end{tabular}
\tablefoot{\tablefoottext{a}{The proportion of the period covered by the daily magnetogram pairs, i.e. the ratio of the number of pairs and the number of days between the first and last pairs.}\tablefoottext{b}{The full width at half maximum (FWHM) of the Gaussian kernel applied to the MDI magnetograms in the comparison to SPM and VSM magnetograms, and to the HMI magnetograms in the comparison to VSM and MDI magnetograms (see Sect. \ref{preparation}).}\tablefoottext{c}{The magnetogram signal threshold applied to the MDI(1) (first column) and SPM data (second column) in the MDI(1)-SPM comparison, and so on (see Sect. \ref{magnetogramnoise}).}}
\end{table*}

HMI magnetograms exhibit 12-h and 24-h periodicities at high magnetogram signal levels \citep{liu12}. This is an instrumental artefact due to the combination of strong Zeeman splitting and the diurnal variation in the line-of-sight (LOS) velocity of the SDO spacecraft. For the HMI-VSM comparisons, for each day, we select the VSM magnetogram taken closest in time to when the LOS velocity of SDO relative to the Sun is at its minimum and the HMI magnetogram taken closest in time to that, similarly for the HMI-MDI comparison.

We exclude from further consideration the daily magnetogram pairs recorded more than three hours apart of one another or which suffer severe instrumental artefacts such as missing values. While MDI and HMI are spaceborne, SPM and VSM are ground-based and therefore subjected to atmospheric seeing. For the comparisons involving SPM and VSM, we also exclude the daily magnetogram pairs where the SPM and VSM magnetograms suffer severe seeing degradation (identified by visual inspection). In Table \ref{instrumentpairs}, we list the number of daily magnetogram pairs and the period they cover in each of the seven instrument comparisons.

\subsection{Data reduction}
\label{preparation}

Our objective is to estimate the uncertainty in measured magnetism due to changes in instrument response by comparing near-simultaneous observations from the various instruments. It is essential to account for factors other than response changes that might affect this comparison. For this reason, we took the following steps to account for the differences in spatial resolution and the disparity between two magnetograms recorded some time apart of one another due to solar rotation. For each instrument comparison, each magnetogram pair, we rotated the solar disc in time to the midpoint of the observation times of the two magnetograms. Then, we resampled the magnetograms from the instrument with the finer pixel scale to the pixel scale of the other instrument. Finally, we smoothed the set of magnetograms with the higher spatial resolution (after the resampling in the previous step) to match that of the other by convolving the former with a Gaussian kernel. The width of the Gaussian kernel, listed in Table \ref{instrumentpairs}, is set at the value that maximises the correlation between the two sets of magnetograms at disc centre.

For the MDI-SPM and MDI-VSM comparisons, the SPM and VSM magnetograms are binned to the pixel scale of MDI and the MDI magnetograms are smoothed to the spatial resolution of the binned SPM and VSM magnetograms. For the HMI-VSM and HMI-MDI comparisons, the HMI magnetograms are binned and smoothed to the pixel scale and spatial resolution of VSM and MDI. The width of the Gaussian kernel differs between the MDI(1)-SPM and MDI(2)-SPM comparisons, between the MDI(2)-VSM(1) and MDI(2)-VSM(2) comparisons, and between the HMI-VSM(2) and HMI-VSM(3) comparisons (Table \ref{instrumentpairs}). This reflects the fact that the spatial resolution of MDI and VSM were altered with the exceptional instrument events discussed in Sect. \ref{instrumentevents}.

\subsection{Magnetogram noise threshold}
\label{magnetogramnoise}

To distinguish magnetic features in the magnetogram data from noise, we counted the points on the solar disc where the absolute magnetogram signal lies above a certain threshold value as magnetic. Let us refer to the magnetogram signal threshold, determined for each of the two sets of magnetograms in each instrument comparison, as the noise threshold. Here and in the rest of the paper, we ignore the points on the solar disc outside the $\mu=0.2$ loci, where $\mu$ denotes the cosine of the heliocentric angle. (The excluded annulus is only about $4\%$ of the solar disc by area.)

It is known for SPM, MDI, VSM, and HMI that the magnetogram noise varies with position on the solar disc \citep{wenzler04,liu12,pietarila13}. For each instrument comparison, each of the two sets of magnetograms, we determined the noise level as a function of disc position following the procedure of \cite{ortiz02}. Taking the magnetogram from a low-activity day \citep[as indicated by the international sunspot number,][]{clette16}, the 1$\sigma$ noise level at each point on the solar disc is given by the standard deviation of the points within a square window centred on that disc position. The window width is set at one-tenth the image width. The result is a map of the 1$\sigma$ noise level. This is repeated for ten other low-activity days. We took the median of the eleven noise level maps at each disc position. This is to minimise any effect of the magnetic activity present in individual magnetograms. The final estimate of the 1$\sigma$ noise level is given by the polynomial surface fit to the resultant median map. Conservatively, we set the noise threshold of the given set of magnetograms at three times the maximum over the solar disc of the 1$\sigma$ noise level map. By setting the noise threshold in this manner, we factor out the variation in noise level with disc position from the subsequent analysis, in particular, the investigation into the effect of the variation in instrument response with disc position on measured magnetism (Sect. \ref{analysisrelationship}).

Let $\nmag$ represent, for a given magnetogram, the proportion of the solar disc where the absolute magnetogram signal is above the noise threshold. Of course, $\nmag$ is systematically higher in the less noisy set of magnetograms in each instrument comparison. (The less noisy a magnetograph, the more magnetic structures present on the solar disc can be distinguished from noise, the greater the value of $\nmag$.) It is important for the purposes of the current study to account for the differences in noise level between the various instruments for the same reason we needed to account for the differences in spatial resolution and observation time (Sect. \ref{preparation}). To this end, we raised the noise threshold of the less noisy set of magnetograms in each instrument comparison such that the mean $\nmag$ matches that in the other set of magnetograms. The final noise threshold is listed in Table \ref{instrumentpairs}. What we have done is to isolate the magnetic activity observable by both instruments in a given comparison by excluding the weak magnetic signals that only the less noisy instrument can distinguish from noise.

By applying the same noise threshold to all the magnetograms from a given instrument in a given comparison, we have implicitly assumed that magnetogram noise is invariant with time. This is a necessary assumption as it is not straightforward to map the noise level separately for each individual magnetogram reliably, at least not by the approach taken here, due to the confounding influence of the magnetic activity present. The noise level of the various instruments might have changed over their operational life. This is at least the case for MDI: \cite{ball11} reported that the noise level of this instrument appears to be higher towards the end of its operational life than at the beginning. To take this into account, for the MDI(2)-SPM comparison, we determined the noise threshold of the MDI magnetograms using observations from close to the end of the period of comparison. We will discuss the implications of assuming magnetogram noise is invariant with time in Sect. \ref{analysisstability}.

\section{Analysis}
\label{analysis}

The aim of this investigation is to address the issue that while various studies have found evidence that the response of solar magnetographs changes with time, magnetogram signal level, and disc position, the uncertainty in measured magnetism due to these instrument fluctuations is not well studied (Sect. \ref{introduction}). There are two main parts to the analysis. In the first part of the analysis (Sect. \ref{analysisstability}), we compare the daily solar disc coverage by magnetic features, $\nmag$ in the two sets of magnetograms in each instrument comparison. From this, we will infer the uncertainty in the longitudinal magnetogram signal, $\blos$ and the disc-integrated photospheric magnetic flux, $\ftot$ associated with the time variation in instrument response. In the second part of the analysis (Sect. \ref{analysisrelationship}), we will derive the conversion factor between the two instruments in each comparison as a function of $\blos$ and distance from disc centre. By looking at the effects of cross-calibrating the two sets of magnetograms with these conversion factors, we will estimate the uncertainty in $\blos$ and $\ftot$ from the variation in instrument response with $\blos$ and distance from disc centre.

By looking at the variation in instrument response with time and with magnetogram signal level and disc position separately, what we are doing is to probe the time variation in instrument response under the assumption that it is invariant with magnetogram signal level and disc position, and vice versa. Of course, in reality, instrument response varies with the various factors simultaneously. However, examining the variation in instrument response with time, magnetogram signal level and disc position simultaneously is made less than straightforward by the fact that photospheric magnetic structures are not evenly distributed in time, field strength or disc position. Active regions and therefore strong magnetic fields are concentrated in periods of high activity and at mid-latitudes \citep[e.g.][]{parnell09}. For this reason, we refrain from such an analysis here.

In their analysis of instrument stability, \cite{arge02} compared the total photospheric magnetic flux and open flux indicated by the various instruments while \cite{riley14} examined the global conversion factor as a function of time (Sect. \ref{introduction}). By the term global we allude to the fact that \cite{riley14} had calculated, for each Carrington rotation, the overall conversion factor between the synoptic magnetograms from two given instruments, neglecting the variation in this quantity with magnetogram signal level and latitude. Due to the greater occurrence of stronger magnetic fields on the solar surface during periods of enhanced activity and the variation in instrument response with magnetogram signal level, the total photospheric magnetic flux, open flux, and global conversion factor can be biased by activity. This is the reason why we depart from these earlier studies and instead examine $\nmag$, which is unlikely to be biased by activity in the same way.

\subsection{Variation with time}
\label{analysisstability}

In this first part of the analysis, for each instrument comparison, we will compare $\nmag$ in the two sets of magnetograms and determine from it the uncertainty in $\blos$ and $\ftot$ associated with the time variation in instrument response. As typical of studies based on longitudinal magnetograms, including \cite{arge02} and \cite{riley14}, we approximate the magnetic flux density within each resolution element by $\blos/\mu$. For a particular magnetogram, $\ftot$ is then given by the integral of $\left|\blos/\mu\right|$ over the solar disc, excluding the image pixels where $\blos$ is below the noise threshold.

Having accounted for the differences in spatial resolution (Sect. \ref{preparation}) and noise level (Sect. \ref{magnetogramnoise}), if two instruments under comparison are stable with time, then $\nmag$ should be, random scatter aside, identical between the two sets of magnetograms. Conversely, instrument changes can produce differences in $\nmag$. For a particular magnetogram, if the instrument response at the time of observation is stronger (weaker) than usual such that magnetogram signals are enhanced (diminished), then $\nmag$ could be biased upwards (downwards). This quantity can also be biased by changes in the magnetogram noise level and the zero level offset. For this investigation, we assume magnetogram noise is stable with time and the zero level offset is negligible, such that we treat any disparity in $\nmag$ to be from the time variation in instrument response alone. We will discuss the implications of both of these assumptions later in this section.

\begin{figure*}
\centering
\includegraphics{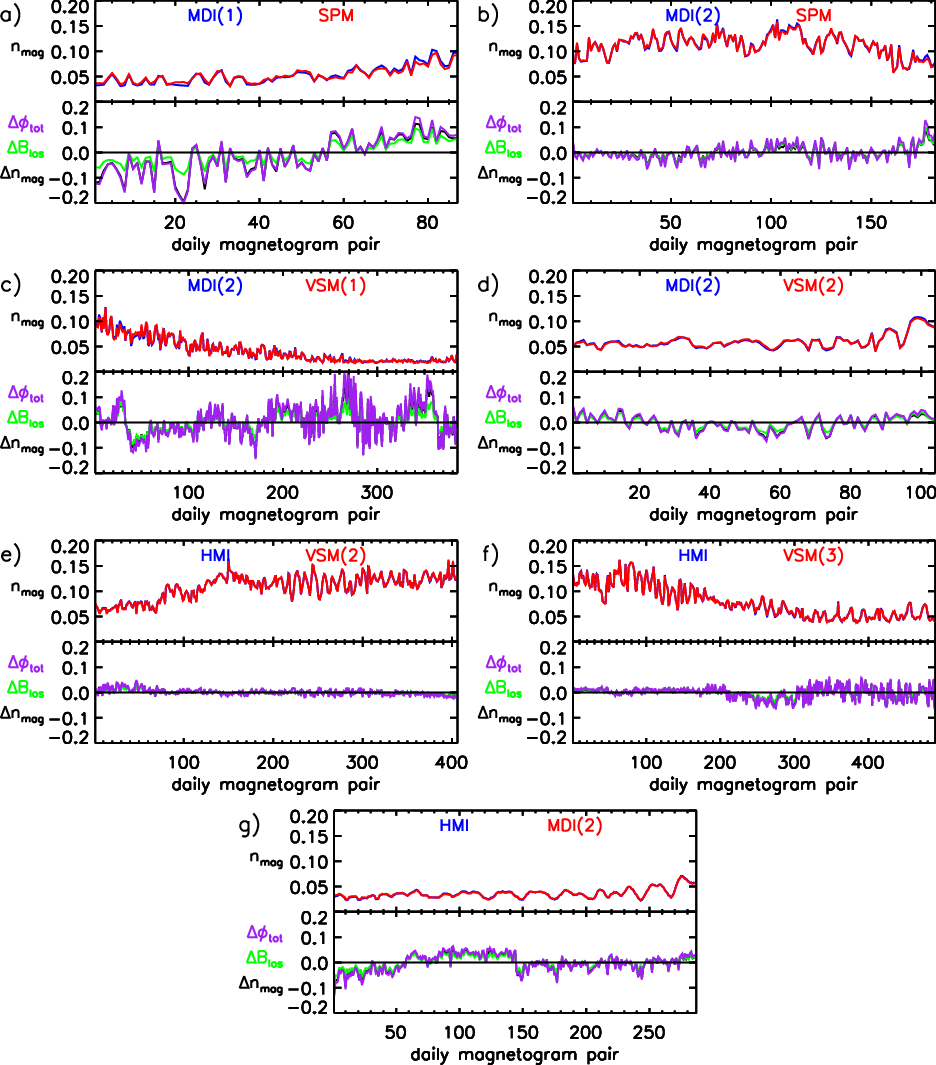}
\caption{As a function of time, how the various magnetographs compare to one another. a) The top plot shows, for the MDI(1)-SPM comparison, the proportion of the solar disc above the magnetogram noise threshold, $\nmag$ in the two sets of magnetograms (blue and red). The bottom plot shows the proportional difference in $\nmag$ between the MDI(1) and SPM magnetograms, $\rdnmag{}_{\rm ,MDI(1)-SPM}$ (black, Equation \ref{rdnmageqn}) and the proportional change in $\blos$ and $\ftot$ in the MDI(1) magnetograms from the time variation in instrument response, $\rdblos{}_{\rm ,MDI(1)-SPM}$ (green, Equation \ref{rdbloseqn}) and $\rdftot{}_{\rm ,MDI(1)-SPM}$ (purple, Equation \ref{rdftoteqn}). b) to g) The corresponding plots for the other instrument comparisons. Due to similarity, the $\rdnmag{}$ and $\rdblos{}$ profiles are largely hidden behind the $\rdftot{}$ profiles. See Sect. \ref{analysisstability}.}
\label{cover}
\end{figure*}

\begin{figure*}
\centering
\includegraphics{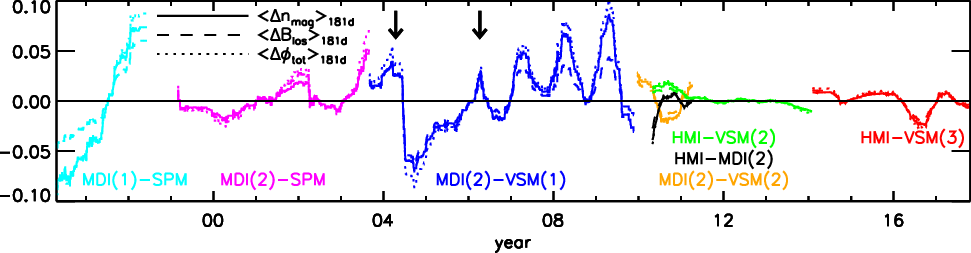}
\caption{The longer-term trend in how the various magnetographs compare to one another. For each instrument comparison, the 181-day running mean of $\rdnmag$ (solid lines), $\rdblos$ (dashed lines), and $\rdftot$ (dotted lines). The arrows mark the relocation of the SOLIS instrument suite in 2004 and the replacement of the VSM modulator in 2006.}
\label{stability}
\end{figure*}

In Fig. \ref{cover}, we chart $\nmag$ in the two sets of magnetograms in each instrument comparison (blue and red) and the relative difference, $\rdnmag$ (black). Taking the MDI(1)-SPM comparison as an example, on day $t$,
\begin{equation}
\label{rdnmageqn}
	\rdnmag{}_{\rm ,MDI(1)-SPM}(t)=2\frac{\nmag{}_{\rm ,MDI(1)}(t)-\nmag{}_{\rm ,SPM}(t)}{\nmag{}_{\rm ,MDI(1)}(t)+\nmag{}_{\rm ,SPM}(t)}.
\end{equation}
Then, we took the 181-day running mean of $\rdnmag$, denoted by $\srdnmag$, to filter out the random scatter from effects such as telescope jitter and short-term temperature fluctuations, drawn in Fig. \ref{stability} (solid lines). For a given instrument comparison, the $\srdnmag$ on a particular day is given by the mean $\rdnmag$ of the daily magnetogram pairs in the 181-day window centred on that day and this is calculated for each day in the period of comparison. For the MDI-SPM comparisons, the daily magnetogram pairs cover only about 10$\%$ of the period of comparison (Table \ref{instrumentpairs}). We set the width of the window at 181 days so that there is sufficient statistics,  i.e. a reasonable number of daily magnetogram pairs in the window. For the other instrument comparisons, even though the magnetogram data cover more of the period of comparison, we kept the window width at 181 days. This is so that we are filtering out the shorter-term variability and isolating the longer-term variability in $\rdnmag$ consistently between the various instrument comparisons. As depicted in Fig. \ref{stability}, the 181-day running mean of $\rdnmag$ reveals this quantity to drift and offset from null. Following the argument in the previous paragraph, we attribute this to the time variation in the relative instrument response. As a test, we repeated this analysis with a window width of 91-day instead. Expectedly, the 91-day running mean retains more of the shorter-term variability in $\rdnmag$ than the 181-day running mean, but the overall trend is largely similar. This analysis is not particularly sensitive to the window width.

Going back to the example of the MDI(1)-SPM comparison, let us assume SPM to be invariant. That is to say, the differences in $\nmag$ between the two sets of magnetograms is from instabilities in the MDI(1) instrument alone. (In the absence of ideal magnetograms, we have to take one of the two instruments in each comparison as the reference.) Let us also suppose that at a particular time, any departure in MDI(1) from the average state of this instrument over the period of comparison rescales the magnetogram signal recorded at that time by a factor, denoted $k_{\rm MDI(1)-SPM}$. For a particular daily magnetogram pair, $k_{\rm MDI(1)-SPM}$ is given by the reciprocal of the factor we have to multiply the MDI(1) magnetogram by such that the difference in $\nmag$ goes to zero. The quantity 
\begin{equation}
\label{rdbloseqn}
	\rdblos{}_{\rm ,MDI(1)-SPM}(t)=k_{\rm MDI(1)-SPM}(t)-1
\end{equation}
represents the proportional change in the MDI(1) magnetogram signal from the time variation in instrument response (green, Fig. \ref{cover}a). The corresponding proportional change in the disc-integrated photospheric magnetic flux is given by
\begin{equation}
\label{rdftoteqn}
	\rdftot{}_{\rm ,MDI(1)-SPM}(t)=\frac{\ftot{}_{\rm ,MDI(1)-SPM}(t)}{\phi^{*}_{\rm tot,MDI(1)-SPM}(t)}-1,
\end{equation}
where $\ftot{}_{\rm ,MDI(1)-SPM}$ and $\phi^{*}_{\rm tot,MDI(1)-SPM}$ denote the disc-integrated photospheric magnetic flux in the MDI(1) magnetogram from day $t$ as it is and after it is rescaled by a factor of $1/k_{\rm MDI(1)-SPM}$ (purple, Fig. \ref{cover}a). This analysis is not sensitive to which instrument we take as the reference in that $\rdblos{}_{\rm ,MDI(1)-SPM}\approx-\rdblos{}_{\rm ,SPM-MDI(1)}$ and $\rdftot{}_{\rm ,MDI(1)-SPM}\approx-\rdftot{}_{\rm ,SPM-MDI(1)}$. The results from the other instrument comparisons are depicted in Figs. \ref{cover}b to \ref{cover}g. As with $\rdnmag$, we applied a 181-day boxcar filter to $\rdblos$ and $\rdftot$ to filter out the random scatter. The result, $\srdblos$ (dashed lines, Fig. \ref{stability}) and $\srdftot$ (dotted lines), trace the drifts and offsets in $\blos$ and $\ftot$ due to the time variation in instrument response.

Looking at Fig. \ref{stability}, with the exception of the MDI(1)-SPM and MDI(2)-VSM(1) comparisons, the various instruments appear relatively stable with respect to one another, with $\srdblos$ and $\srdftot$ undulating about null and well below $5\%$ almost everywhere. In the MDI(1)-SPM comparison (cyan), $\srdblos$ and $\srdftot$ drifted monotonically, by about $10\%$, over the period of comparison. The MDI(2)-SPM comparison (pink) does however indicate that the two instruments are relatively stable with respect to one another after the SoHO vacation and MDI refocus. In the MDI(2)-VSM(1) comparison (blue), $\srdblos$ and $\srdftot$ dropped sharply by about $10\%$ in 2004 and rose gradually by a similar margin over the next two years before they started to exhibit an annual modulation of about $5\%$ to $10\%$. We attribute the drop in 2004 and the drift over the next two years to the relocation of the SOLIS instrument suite in 2004 and the VSM modulator degrading gradually between 2003 and 2006 before being replaced \citep{pietarila13,riley14}, marked in Fig. \ref{stability} by the arrows. The strong annual modulation could allude to unaccounted annual temperature fluctuations.

The root-mean-square (RMS) value of $\srdblos$ and $\srdftot$, denoted $\rmsblost$ and $\rmsftott$, is an estimate of the proportional uncertainty in $\blos$ and $\ftot$ from the time variation in instrument response, tabulated in Table \ref{uncertainty}. Excluding the MDI(1)-SPM and MDI(2)-VSM(1) comparisons, the uncertainty in the observed LOS magnetic flux density and disc-integrated photospheric magnetic flux does not exceed $2\%$: $\rmsblost$ ranged from $0.6\%$ to $1.6\%$ and $\rmsftott$ from $0.8\%$ to $1.9\%$. Of course, the exact value of $\rmsblost$ and $\rmsftott$, calculated from the 181-day running mean of $\rdblos$ and $\rdftot$, depends on the width of the boxcar filter. The narrower (wider) the boxcar filter, the less (more) shorter-term variability is filtered out and the higher (lower) the value of $\rmsblost$ and $\rmsftott$. For example, if we had taken the 91-day running mean instead, $\rmsblost$ and $\rmsftott$ would be a fraction or few tens of percent higher. Nonetheless, $\rmsblost$ and $\rmsftott$ remain useful as indications of the scale of the uncertainties in $\blos$ and $\ftot$ due to to instrument instabilities and how the various instrument comparisons match up to one another in this regard.

\begin{table}
\caption{The uncertainty in measured solar magnetism due to the time variation in instrument response.}
\label{uncertainty}
\centering
\begin{tabular}{ccc}
\hline\hline
 & $\rmsblost$ & $\rmsftott$ \\
Comparison & $\left[\%\right]$ & $\left[\%\right]$ \\
\hline
MDI(1)-SPM    & 3.56  & 6.15  \\
MDI(2)-SPM    & 1.57  & 1.94  \\
MDI(2)-VSM(1) & 2.47  & 4.17  \\
MDI(2)-VSM(2) & 1.18  & 1.73  \\
HMI-VSM(2)    & 0.561 & 0.813 \\
HMI-VSM(3)    & 0.807 & 1.10  \\
HMI-MDI(2)    & 0.733 & 1.15  \\
\hline
\end{tabular}
\tablefoot{For each instrument-instrument comparison, the proportional uncertainty in the LOS magnetic flux density, $\rmsblost$ and disc-integrated photospheric magnetic flux, $\rmsftott$ due to the time variation in instrument response. The values are scaled by a factor of 100 to express them as percentages. See Sect. \ref{analysisstability} for details.}
\end{table}

\begin{figure*}
\centering
\includegraphics{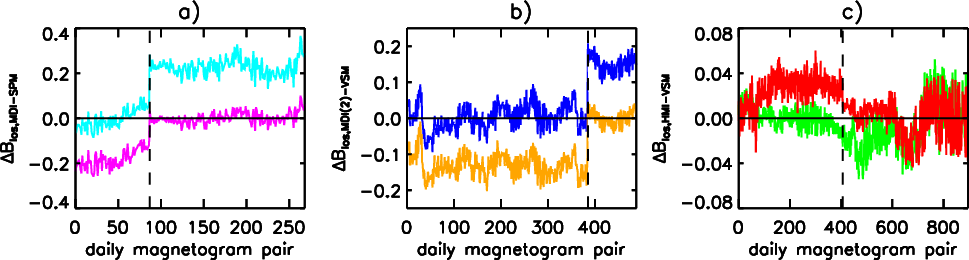}
\caption{The proportional change in magnetogram signal due to the time variation in instrument response. a) Cyan: The proportional change in the MDI magnetogram signal, relative to SPM, from the time variation in instrument response, $\rdblos{}_{\rm ,MDI-SPM}$ derived combining the MDI(1)-SPM and MDI(2)-SPM data sets and applying the procedure of the MDI(1)-SPM comparison. Pink: The result from applying the procedure of the MDI(2)-SPM comparison to the combined data set. b) The estimates of $\rdblos{}_{\rm ,MDI(2)-VSM}$ from replicating the procedure of the MDI(2)-VSM(1) (blue) and MDI(2)-VSM(2) comparisons (orange) on the combination of the data sets from these two comparisons. c) The estimates of $\rdblos{}_{\rm ,HMI-VSM}$ from replicating the procedure of the HMI-VSM(2) (green) and HMI-VSM(3) comparisons (red) on the combination of the data sets from these two comparisons. The colour-coding follows Fig. \ref{stability}. The dashed lines mark a) the SoHO vacation and MDI refocus and b) the VSM CCD camera upgrade and c) filling gas switch. See Sects. \ref{instrumentevents} and \ref{analysisstability}.}
\label{exceptional}
\end{figure*}

The spatial resolution of MDI and VSM had changed with the exceptional instrument events discussed in Sect. \ref{instrumentevents}, which prompted us to distinguish between MDI and VSM observations from before and after each event in the analysis. To elucidate the effect the SoHO vacation and MDI refocus might have had on the response of this instrument, we combined the MDI(1)-SPM and MDI(2)-SPM data sets and applied to it the procedure of the MDI(1)-SPM comparison. We depict the proportional change in the MDI magnetogram signal due to the time variation in instrument response, $\rdblos{}_{\rm ,MDI-SPM}$ this yields in Fig. \ref{exceptional}a (cyan). We also show the result of applying the procedure of the MDI(2)-SPM comparison to the combined data set (pink). If the relative instrument response did not change with the SoHO vacation and MDI refocus, then $\rdblos{}_{\rm ,MDI-SPM}$ should be at similar levels before and after this event (dashed line). The SoHO vacation and MDI refocus is accompanied by a $15\pm2\%$ offset in $\rdblos{}_{\rm ,MDI-SPM}$. The MDI magnetograms from before and after this event can be approximately reconciled to one another by scaling the former by a factor of 1.15 or the latter by the reciprocal, 0.87. The implicit assumption here is that SPM is stable over the period of the event, such that the offset in $\rdblos{}_{\rm ,MDI-SPM}$ is entirely due to changes in the response of MDI.

In the same manner, we combined the MDI(2)-VSM(1) and MDI(2)-VSM(2) data sets to examine the effect of the VSM CCD camera upgrade (Fig. \ref{exceptional}b) and the HMI-VSM(2) and HMI-VSM(3) data sets to examine the effect of the VSM filling gas switch (Fig. \ref{exceptional}c). The CCD camera upgrade is accompanied by a $18\pm2\%$ offset in $\rdblos{}_{\rm ,MDI-VSM}$ (Fig. \ref{exceptional}b). Assuming the response of MDI did not change at the same time, pre-upgrade and post-upgrade VSM magnetograms can be reconciled to one another by scaling the latter by a factor of 1.18 or the former by the reciprocal, 0.85. The change in $\rdblos{}_{\rm ,HMI-VSM}$ at the point of the filling gas switch is somewhat obscured by the fluctuations in this quantity at other times (Fig. \ref{exceptional}c). The degradation in spatial resolution caused by the filling gas switch does not appear to have a greater effect on the response of VSM than the instabilities seen in its regular operation. The results obtained for $\rdftot$ are similar to what we have shown here for $\rdblos$, omitted to avoid repetition.

Next, let us examine the sum effect of the gradual drift in the response of MDI relative to SPM prior to the SoHO vacation and MDI refocus (i.e. what we have noted of the MDI(1)-SPM comparison) and of this event itself. Looking at Fig. \ref{exceptional}a, from when MDI started operation to the refocus of this instrument (dashed line), $\rdblos{}_{\rm ,MDI-SPM}$ rose by $0.24\pm0.02$ (the sum of the gradual upward drift and the $0.15\pm0.02\%$ offset after the SoHO vacation and the MDI refocus). In other words, the MDI magnetogram signal, relative to the SPM magnetogram signal, increased by $24\pm2\%$. Of course, this variation in relative response is likely from changes to both instruments and not either one alone. Nonetheless, the possibility that the response of MDI might have changed by any amount close to the rather stark figure of $24\pm2\%$ is consequential. The MDI instrument operated from 1996 to 2011, a period which encompasses solar cycle 23 in its entirety. It remains the only solar telescope to have recorded seeing-free full-disc magnetograms over a complete solar cycle from minimum to minimum. For this reason, it is a particularly useful data set to examine how various aspects of the photospheric magnetic field changes over the solar cycle, such as ephemeral regions \citep{hagenaar03}, the magnetic network \citep{jin14} and the internetwork magnetic field \citep{meunier18}. The observation here that the response of MDI could have increased significantly over the period of 1996 to 1999, incidentally the ascending phase of solar cycle 23, warrants caution when interpreting the results from such studies and should be taken into account in future investigations. For example, \cite{meunier18} found the strength of the internetwork magnetic field, as apparent in MDI observations, to vary in phase with the solar cycle. However, the rise over the ascending phase of solar cycle 23 (1996 to 1999) is similar to the change in the response of MDI relative to SPM over the same period. Consequently, it cannot be ruled out that this part of their result might, at least in part, be an instrument artefact. We discuss this in detail in Section \ref{examplesm18}.

In the analysis presented in this section, we had assumed that magnetogram noise is stable with time and the zero level offset is negligible. Recall from Sect. \ref{magnetogramnoise}, the assumption that magnetogram noise is constant is almost unavoidable because it is not straightforward to trace its time variation. The apparent effect of the time variation in instrument response on $\blos$ and $\ftot$ revealed by the above analysis might contain contributions by noise fluctuations. The influence of instrument instability could therefore be weaker than estimated here. The residual zero level offset of the various instruments examined in this study, i.e. what is not already corrected for by the respective instrument teams, is relatively small \citep[well below 1 G,][]{getachew19}. Also, since the zero level offset would have the equal and opposite effect on positive and negative magnetic signals, we expect the effect of the bias in positive and in negative magnetic signals on $\rdnmag$, $\rdblos$, and $\rdftot$ to at least partially cancel out. Taking these factors into consideration, any effect the zero level offset might have had on the above analysis is likely minute.

It is less than straightforward to compare the results we have obtained here to what \cite{riley14}, who had also examine the stability of SPM, MDI, VSM, and HMI by comparing them to one another, had reported. While we had inferred the time variation in instrument response from the surface coverage by magnetic features in full-disc magnetograms, \cite{riley14} looked at the global conversion factor based on magnetic synoptic maps. As noted in the beginning of Sect. \ref{analysis}, the global conversion factor can be biased by activity. Since the process of constructing synoptic magnetograms from full-disc magnetograms evidently involves modifications to the latter, and the various instrument teams can differ in how this is done, the synoptic magnetograms from the various instruments might not compare to one another in the same way as their full-disc magnetograms. A direct comparison is also made difficult by differences in how the two studies organised the magnetogram data. We split the MDI and VSM data sets, distinguishing between observations from before and after the exceptional instrument events detailed in Sect. \ref{exceptional}, while \cite{riley14} combined the SPM and VSM data sets into one, which they termed `SOLIS'. While the \cite{riley14} analysis did indicate fluctuations in how MDI and `SOLIS' compare with one another, see Fig. 9 in their paper, given the factors we have just discussed, a direct comparison with what we have obtained in this study is difficult.

\subsection{Variation with magnetogram signal level and disc position}
\label{analysisrelationship}

In this second part of the study, we examine the effect of the variation in instrument response with magnetogram signal level and disc position on measured magnetism. First, we determine the conversion factor between the various instruments as a function of the absolute magnetogram signal level, $\ablos$ and the cosine of the heliocentric angle, $\mu$ and cross-calibrate the observations from the different instruments with the result. We neglect the non-radial component of the variation with disc position, a step we will justify later in this discussion. Then, we repeat this analysis with the global conversion factor, the overall ratio of the magnetogram signal registered by two given instruments. Finally, by comparing the results from the two analyses, we estimate the uncertainty in $\blos$ and $\ftot$ due to the variation in instrument response with magnetogram signal level and distance from disc centre. It is worth emphasising that while earlier investigations have examined the conversion factor between the various instruments (Sect. \ref{introduction}), this is the first study of its kind to probe the associated uncertainty in measured magnetism in detail. As in the first part of the study and for similar reasons (Sect. \ref{analysisstability}), we assume the zero level offset is negligible.

For each instrument comparison, we determine the conversion factor as a function of $\ablos$ and $\mu$ adopting the histogram equalisation (HE) method \citep{jones01}. The idea behind HE is, rather than comparing the magnetogram signal in simultaneous observations from the two instruments in question directly, to instead bin the points in the observations from each instrument by the magnetogram signal and compare the bin-average magnetogram signal. The procedure here broadly follows \cite{yeo14}. Let us take the MDI(1)-SPM comparison as an example. We divide the solar disc by $\mu$ into 10 intervals, defined such that the number of magnetic image pixels, i.e. image pixels above the noise threshold (Table \ref{instrumentpairs}), in each interval is roughly even. The interval at disc centre is $0.97<\mu\leq1.00$. Going from disc centre to limb, the interval width increases steadily from 0.03 to 0.35, such that the interval closest to the limb is $0.20<\mu\leq0.55$.  We collate the magnetic image pixels present in the two sets of magnetograms. Suppose a particular SPM magnetogram and the corresponding MDI(1) magnetogram have, in a given $\mu$-interval, $M$ and $N$ magnetic image pixels, respectively. If $M<N$, we isolate the $M$ magnetic pixels in the SPM magnetogram and just the strongest $M$ of the $N$ magnetic pixels in the MDI(1) magnetogram, and vice versa if $M>N$. This is to factor out the imbalance between $M$ and $N$ due to factors such as the time variation in instrument response and the difference in observation time. For each instrument and interval of $\mu$, we bin the magnetic image pixels collated from all the magnetograms by $\ablos$ into 2000 bins of equal population and calculate the average $\ablos$ within each bin. This yields, for each interval of $\mu$, a bin-averaged $\left|B_{\rm los,MDI(1)}\right|$ series and a bin-averaged $\left|B_{\rm los,SPM}\right|$ series that together describe the relationship between MDI(1) and SPM magnetogram signals at this distance from disc centre. Let us refer to this as the HE table.

\begin{figure}
\centering
\includegraphics{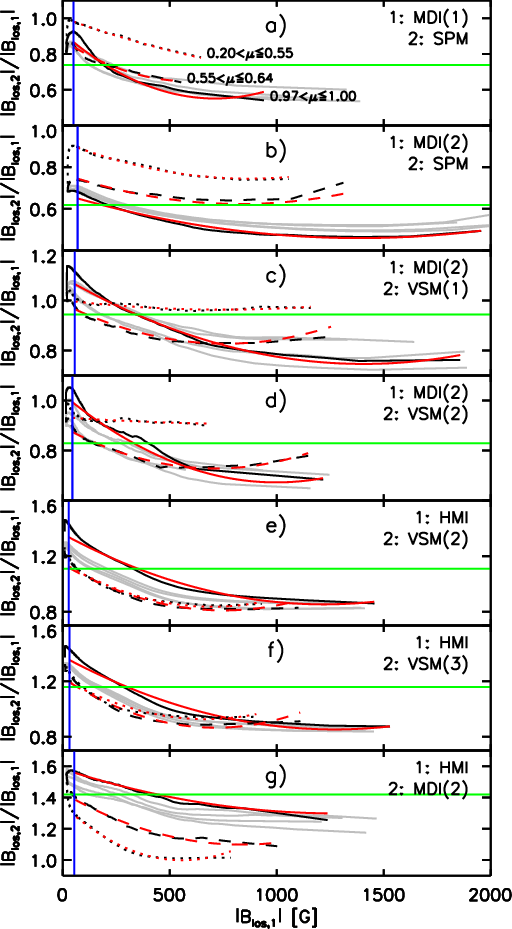}
\caption{The conversion factor between the various magnetographs. a) Black: The MDI(1)-to-SPM conversion factor as a function of the MDI(1) magnetogram signal, $\left|B_{\rm los,MDI(1)}\right|$ at $0.97<\mu\leq1.00$ (solid curves), $0.55<\mu\leq0.64$ (dashed curves), and $0.20<\mu\leq0.55$ (dotted curves). Red: The corresponding bivariate polynomial in $\left|B_{\rm los,MDI(1)}\right|$ and $\mu$ fit (Equation \ref{bicubicfiteqn}). The points below three times the magnetogram noise threshold (blue line) were excluded from the fit. Green: The global conversion factor (Equation \ref{lineareqn}). Grey: The conversion factor-versus-$\left|B_{\rm los,MDI(1)}\right|$ profile in each of the four quadrants of the solar disc. b) to g) The corresponding plots for the other instrument comparisons. See Sect. \ref{analysisrelationship} for details.}
\label{relationship}
\end{figure}

In Fig. \ref{relationship}a, we chart the MDI(1)-to-SPM conversion factor, $\left|B_{\rm los,SPM}\right|/\left|B_{\rm los,MDI(1)}\right|$ in the $0.97<\mu\leq1.00$ (black solid line), $0.55<\mu\leq0.64$ (black dashed line), and $0.20<\mu\leq0.55$ intervals (black dotted line) as a function of $\left|B_{\rm los,MDI(1)}\right|$. Going from high to low $\left|B_{\rm los,MDI(1)}\right|$, the MDI(1)-to-SPM conversion factor rises gradually up to $\left|B_{\rm los,MDI(1)}\right|$ of about 50 G before dropping rapidly at lower $\left|B_{\rm los,MDI(1)}\right|$ instead. In Figs. \ref{relationship}b to \ref{relationship}g, we depict the results from the other instrument comparisons. As with the MDI(1)-SPM comparison, in each instance, going towards the origin, there is an abrupt change in the trend in the conversion factor with $\ablos$. This was similarly noted by \cite{yeo14}, who argued that at low magnetogram signal levels, the conversion factor might be biased by noise such that it no longer indicates the true relationship between the magnetogram signal scale of the two instruments in question.

Taking the HE table from the MDI(1)-SPM comparison, we fit a function of the form
\begin{equation}
\label{bicubicfiteqn}
	\left|B_{\rm los,SPM}\right|=\sum^{3}_{i=1}\sum^{3}_{j=0}c_{ij{\rm,MDI(1)-SPM}}\left|B_{\rm los,MDI(1)}\right|^{i}\mu^{j},
\end{equation}
where $c_{ij{\rm,MDI(1)-SPM}}$ are fit parameters (red curves, Fig. \ref{relationship}a). As the HE table might be biased by magnetogram noise at low $\ablos$, noted in the previous paragraph, we exclude the points in the HE table where $\left|B_{\rm los,MDI(1)}\right|$ is below three times the MDI(1) noise threshold (blue line) from the regression. What we have done here is to capture the information in the HE table, which is essentially a lookup table of how MDI(1) and SPM magnetogram signals relate to one another at each of the 10 intervals of $\mu$, in a relationship that describes the SPM magnetogram signal as a function of the MDI(1) magnetogram signal and distance from disc centre. Taking this relationship, we rescale the MDI(1) magnetograms to the SPM magnetogram signal scale. In the same manner, we cross-calibrate the two sets of magnetograms in each of the other six instrument comparisons. To aid the following discussion, let us refer to this analysis as case A.

Next, we repeat the above analysis with the global conversion factor. Going back to the example of the MDI(1)-SPM comparison, we derive the HE table as before, except taking the entire solar disc as a whole. To the scatter plot of the bin-average $\left|B_{\rm los,SPM}\right|$ and $\left|B_{\rm los,MDI(1)}\right|$ values in the HE table, we fit a straight line that is constrained to pass through the origin. In other words, we assume
\begin{equation}
\label{lineareqn}
	\left|B_{\rm los,SPM}\right|=m_{\blos{\rm,MDI(1)-SPM}}\left|B_{\rm los,MDI(1)}\right|,
\end{equation}
where the fit parameter $m_{\blos{\rm,MDI(1)-SPM}}$ indicates the global MDI(1)-to-SPM conversion factor (green line, Fig. \ref{relationship}a). For the same reason as in case A, we exclude the points in the HE table where $\left|B_{\rm los,MDI(1)}\right|$ is below three times the MDI(1) noise threshold from the regression. Finally, we rescaled the MDI(1) magnetograms by $m_{\blos{\rm,MDI(1)-SPM}}$. Let us refer to this analysis as case B.

Various studies have examined the conversion factor between the same instruments we are looking at here \citep{jones01,wenzler04,demidov08,liu12,pietarila13,riley14,yeo14}. The objective of the current study is not to investigate the conversion factor, but rather to use this information to estimate the effect of the variation in instrument response with magnetogram signal level and distance from disc centre on measured magnetism. In Sect. \ref{cfcomparison}, we will compare the conversion factors derived here with what was reported in the earlier efforts. Otherwise, we refer the reader to the cited studies for more detailed discussions of the conversion factor between the various instruments, including the instrumental differences that underlie the observed values.

Let us examine the effect of the cross-calibration. For each instrument comparison, taking the HE table derived in case A by binning the points in the magnetogram observations by $\ablos$ and by $\mu$ and taking the bin-average, we calculate the RMS of the relative difference between the $\ablos$ registered by the two instruments, $\rmsblos$. Taking the example of the MDI(1)-SPM comparison, this is given by
\begin{equation}
\sigma_{\blos{\rm,MDI(1)-SPM}}=\sqrt{
\frac{1}{n}\sum\left(2\frac{\left|B_{\rm los,MDI(1)}\right|-\left|B_{\rm los,SPM}\right|}{\left|B_{\rm los,MDI(1)}\right|+\left|B_{\rm los,SPM}\right|}\right)^2},
\end{equation}
where the summation is over the $n=20000$ bin-average values in the HE table. The quantity $\rmsblos$ is an estimate of the proportional uncertainty in $\blos$ due to the systematic differences between the two sets of magnetograms under comparison. In a similar manner, we estimate the proportional uncertainty in $\ftot$, $\rmsftot$ by calculating the RMS of the relative difference between $\ftot$ in the two sets of magnetograms in each instrument comparison. This is repeated for the cross-calibrated data from case A and case B. Let us denote the result of repeating this computation with the cross-calibrated data from case A as $\rmsblosa$ and $\rmsftota$ and likewise that from case B as $\rmsblosb$ and $\rmsftotb$. The results are tabulated in Table \ref{agreement}. We note here that this analysis is commutative in that the results from switching the roles of the two instruments in each comparison, e.g. rescaling SPM magnetograms to the MDI magnetogram signal scale instead of vice versa, are similar.

\begin{table}
\caption{The uncertainty in measured solar magnetism due to the systematic differences between the various magnetographs.}
\label{agreement}
\centering
\begin{tabular}{ccccc}
\hline\hline
 & $\rmsblos$ & $\rmsblosa$ & $\rmsblosb$ & $\rmsblosmm$ \\
Comparison & $\left[\%\right]$ & $\left[\%\right]$ & $\left[\%\right]$ & $\left[\%\right]$ \\
\hline
MDI(1)-SPM    & 40.7 & 2.39 & 8.08 & 7.71 \\
MDI(2)-SPM    & 23.5 & 5.28 & 14.4 & 13.4 \\
MDI(2)-VSM(1) & 5.51 & 2.87 & 9.37 & 8.92 \\
MDI(2)-VSM(2) & 39.7 & 2.52 & 13.9 & 13.7 \\
HMI-VSM(2)    & 26.2 & 5.19 & 12.6 & 11.5 \\
HMI-VSM(3)    & 10.8 & 3.59 & 11.8 & 11.3 \\
HMI-MDI(2)    & 19.6 & 4.36 & 14.0 & 13.3 \\
\hline
 & $\rmsftot$ & $\rmsftota$ & $\rmsftotb$ & $\rmsftotmm$ \\
Comparison & $\left[\%\right]$ & $\left[\%\right]$ & $\left[\%\right]$ & $\left[\%\right]$ \\
\hline
MDI(1)-SPM    & 20.2 & 7.97 & 24.8 & 23.4 \\
MDI(2)-SPM    & 39.9 & 5.00 & 12.1 & 11.0 \\
MDI(2)-VSM(1) & 7.91 & 8.13 & 15.2 & 12.9 \\
MDI(2)-VSM(2) & 13.6 & 5.81 & 13.0 & 11.6 \\
HMI-VSM(2)    & 14.2 & 5.68 & 9.18 & 7.21 \\
HMI-VSM(3)    & 21.5 & 8.76 & 14.3 & 11.3 \\
HMI-MDI(2)    & 36.6 & 5.01 & 8.62 & 7.02 \\
\hline
\end{tabular}
\tablefoot{Top: For each instrument comparison, the proportional uncertainty in $\blos$ in the original data, $\rmsblos$ and in the cross-calibrated data from case A, $\rmsblosa$ and case B, $\rmsblosb$. The proportional uncertainty in $\blos$ due to the variation in instrument response with magnetogram signal level and distance from disc centre, $\rmsblosmm$ is also listed. Bottom: The same for the proportional uncertainty in the disc-integrated photospheric magnetic flux, $\rmsftot$. The values are scaled by a factor of 100 to express them as percentages. See Sect. \ref{analysisrelationship} for details.}
\end{table}

Looking at Table \ref{agreement}, the case B cross-calibration brought $\blos$ and $\ftot$ in the two sets of magnetograms in each instrument comparison closer together, i.e. $\rmsblosb<\rmsblos$ and $\rmsftotb<\rmsftot$, in every instance except the following. For the MDI(2)-VSM(1) and HMI-VSM(3) comparisons, $\rmsblosb>\rmsblos$ and for the MDI(1)-SPM and MDI(2)-VSM(1) comparisons, $\rmsftotb>\rmsftot$. The case A cross-calibration brought $\blos$ and $\ftot$ in the two sets of magnetograms in each instrument comparison closer together except in the MDI(2)-VSM(1) comparison, where $\rmsftota>\rmsftot$. Comparing case A and case B, without exception, $\rmsblosa<\rmsblosb$ and $\rmsftota<\rmsftotb$. As was to be expected, accounting for the rather significant variation in the conversion factor with $\ablos$ and $\mu$ (Fig. \ref{relationship}) in the cross-calibration (case A) reconciles the observations from the various instruments more closely than if it were ignored (case B). This limitation of the case B analysis and the observation in Sect. \ref{analysisstability} that the MDI(1)-SPM and MDI(2)-VSM(1) comparisons are particularly affected by instrument instabilities are likely, at least in part, why $\rmsblos$ and $\rmsftot$ are higher in the cross-calibrated data than in the original data in the instances we have highlighted above.

The proportional uncertainty in $\blos$ in the original data, $\rmsblos$, is the result of all the factors that contributed to the systematic differences between the two sets of magnetograms under comparison. Let $\sigma_{\blos,1}$, $\sigma_{\blos,2}$, $\sigma_{\blos,3}$, and so on represent the proportional uncertainty in $\blos$ due to individual factors. Let us suppose that the different contributions to the uncertainty in $\blos$ combine in quadrature, such that
\begin{equation}
\label{quadrature1}
    \rmsblos^2=\sigma_{\blos,1}^2+\sigma_{\blos,2}^2+\sigma_{\blos,3}^2+\ldots
\end{equation}
Cross-calibration can be thought of as the process of accounting for certain factors, removing the terms corresponding to them from the right hand side of Equation \ref{quadrature1}. The square root of the sum of the remaining terms, which correspond to the factors that remain unaccounted for, yields $\rmsblos$ in the cross-calibrated data. In case B, by the cross-calibration with the global conversion factor, we accounted for the differences in absolute scale between the various instruments. Following the argument we had just presented,
\begin{equation}
\label{quadrature2}
    \rmsblos^2=\rmsblosb^2+\rmsblosabs^2,
\end{equation}
where $\rmsblosabs$ represents the proportional uncertainty in $\blos$ associated with the uncertainty in the absolute scale. In case A, by determining the conversion factor as a function of $\ablos$ and $\mu$, we additionally accounted for the variation in instrument response with magnetogram signal level and distance from disc centre, such that
\begin{equation}
\label{quadrature3}
    \rmsblos^2=\rmsblosa^2+\rmsblosabs^2+\rmsblosmm^2,
\end{equation}
where $\rmsblosmm$ represents the proportional uncertainty in $\blos$ due to this factor. Combining Equations \ref{quadrature2} and \ref{quadrature3},
\begin{equation}
\label{quadrature4}
    \rmsblosmm=\sqrt{\rmsblosb^2-\rmsblosa^2}.
\end{equation}
By the same argument, the proportional uncertainty in $\ftot$ due to the variation in instrument response with magnetogram signal level and distance from disc centre, $\rmsftotmm$ is given by
\begin{equation}
\label{quadrature5}
    \rmsftotmm=\sqrt{\rmsftotb^2-\rmsftota^2}.
\end{equation}
Taking Equations \ref{quadrature4} and \ref{quadrature5}, we calculate $\rmsblosmm$ and $\rmsftotmm$, tabulated in Table \ref{agreement}.

The variation in instrument response with magnetogram signal level and distance from disc centre produced an overall uncertainty of 8$\%$ to 14$\%$ in $\blos$ and of 7$\%$ to 23 $\%$ in $\ftot$ ($\rmsblosmm$ and $\rmsftotmm$, Table \ref{agreement}). This rather significant uncertainty is why the cross-calibration with the global conversion factor, which neglects this variability, only reconciled the $\blos$ registered by the various instruments to within about 5$\%$ to 9$\%$ and $\ftot$ to 9$\%$ to 25$\%$ ($\rmsblosb$ and $\rmsftotb$). Accounting for the variation in instrument response with magnetogram signal level and distance from disc centre enabled us to reconcile $\blos$ to within 2$\%$ to 5$\%$ and $\ftot$ to within 5$\%$ to 9$\%$  ($\rmsblosa$ and $\rmsftota$), demonstrating the importance in taking it into account when comparing observations from multiple solar magnetographs. For each instrument comparison, both $\rmsblosa$ and the uncertainty in $\blos$ due to the time variation in instrument response, $\rmsblost$ (Table \ref{uncertainty}) are smaller than $\rmsblosmm$, and likewise for $\ftot$. This suggests that the uncertainty in measured magnetism and the systematic differences between the observations from the various instruments is dominated by the variation in instrument response with magnetogram signal level and distance from disc centre.

We had confined this analysis to the radial component of the variation in instrument response with disc position even though the instrument response at a given distance from disc centre might also vary with polar angle. It is less than straightforward to determine the conversion factor as a function of $\ablos$, $\mu$, and polar angle simultaneously. One obstacle is the fact that higher magnetogram signals are concentrated at mid-latitudes. However, as we will demonstrate next, the variation in instrument response with polar angle is likely small compared to the variation with magnetogram signal level and distance from disc centre, such that its omission here is not detrimental to the objectives of the study.

In the case A analysis, we had divided the solar disc by $\mu$ into 10 intervals and derived the conversion factor as a function of $\ablos$ within each interval (black curves, Fig. \ref{relationship}). We repeat this computation, except now we divide the solar disc by the central meridian and the east-west line crossing the disc centre into four quadrants instead. The conversion factor-versus-$\ablos$ profile within each quadrant are drawn in grey, alongside the earlier results. The offset between the various profiles is due to the quadrant-to-quadrant differences in instrument response and distribution of magnetic image pixels with $\mu$. As is visibly evident, for the various instrument comparisons, this offset is relatively small as compared to the variation in the conversion factor with $\ablos$ and $\mu$. Within the limits of this analysis, there is little evidence that the variation in instrument response with polar angle might be significant.

\section{Discussion}
\label{discussion}

\subsection{Comparison between the conversion factors from the current study and earlier studies}
\label{cfcomparison}

In Sect. \ref{analysisrelationship}, for the purpose of quantifying the uncertainty in measured magnetism due to the variation in instrument response with magnetogram signal level and distance from disc centre, we derived, for each instrument comparison, the conversion factor as a function of $\blos$ and $\mu$ (black curves, Fig. \ref{relationship}) and also the global conversion factor (green lines). In this section, we discuss the earlier studies that had examined the conversion factor between the same instruments we are looking at here \citep{jones01,wenzler04,demidov08,liu12,pietarila13,riley14,yeo14}.

\cite{jones01}, \cite{wenzler04}, and \cite{demidov08} examined the conversion factor between SPM and MDI. \cite{jones01} found a value of 0.7 for the MDI(2)-to-SPM global conversion factor. \cite{demidov08} found a value of 0.9 for the MDI(2)-to-SPM global conversion factor. They also determined the MDI(2)-to-SPM conversion factor as a function of $\blos$: going from low to high $\blos$, the conversion factor declines from 1.2 to 0.8. (We have divided the values reported by the authors by a factor of 1.46 to account for the fact that they had scaled the SPM magnetograms by this factor prior to analysis.)

In this study, we retrieved a value of 0.6 for the MDI(2)-to-SPM global conversion factor (green line, Fig. \ref{relationship}b). As for the conversion factor determined as a function of $\blos$ and $\mu$, going from low to high $\blos$, it decreases from 0.7 to 0.5 around disc centre (black solid line) and from 0.9 to 0.8 near the limb (black dotted line). The values we have obtained are a fraction lower than what \cite{jones01} and \cite{demidov08} had reported, likely due to the following. We had resampled the SPM magnetograms to the MDI pixel scale and smoothed the MDI magnetograms with a Gaussian kernel to account for the difference in spatial resolution (Sect. \ref{preparation}). \cite{jones01} and \cite{demidov08} had similarly resampled the SPM magnetograms but kept the MDI magnetograms as they are, therefore only partially accounted for the difference in spatial resolution. We cannot make a direct comparison with the \cite{wenzler04} study. In their analysis, the authors had combined MDI observations from before and after the SoHO vacation and the refocus of this instrument. Also, they had examined the SPM and MDI data at their original spatial resolution.

Counting image pixels with $\blos$ below and above a certain threshold level as weak and strong fields, \cite{liu12} calculated the HMI-to-MDI(2) conversion factor for strong and weak fields separately and within various intervals of $\mu$, returning values ranging from 1.1 to 1.5 (see Tables 1 and 2 in their study). The conversion factor is higher around disc centre and for weak fields than near the limb and for strong fields. In a similar analysis, \cite{pietarila13} derived MDI(2)-to-VSM(1) conversion factors of 0.8 and 1.0 for strong and weak fields, respectively, and HMI-to-VSM(2) conversion factors of 0.7 and 1.0 for the same categories. 

From this study, the HMI-to-MDI(2) conversion factor, determined as a function of $\blos$ and $\mu$, ranges from 1.0 to 1.6 (Fig. \ref{relationship}g), which encompasses the range of values reported by \cite{liu12}. Importantly, consistent with this earlier study, the conversion factor decreases from disc centre to limb and from low to high $\blos$. The MDI(2)-to-VSM(1) conversion factor ranges from 0.8 to 1.1 (Fig. \ref{relationship}c) and the HMI-to-VSM(2) conversion factor from 0.8 to 1.4 (Fig. \ref{relationship}e), broadly consistent with the values reported by \cite{pietarila13}. In both instances, the conversion factor increases with decreasing $\blos$, similar to this earlier study. Keeping in mind that \cite{liu12} and \cite{pietarila13} had partitioned the magnetogram data into broader intervals of $\blos$ and $\mu$ than what we have done here, a close match to the current results is not expected.

We cannot make a direct comparison with \cite{riley14} and \cite{yeo14}. Comparing the conversion factors we have derived as part of the analysis presented in Sect. \ref{analysisrelationship} to what \cite{riley14} reported is less than straightforward for the same reasons why we could not compare the analysis presented in Sect. \ref{analysisstability} to this earlier study, discussed in that section. Similar to what we have done here, \cite{yeo14} derived the conversion factor between SPM, MDI, and HMI as a function of $\blos$ and $\mu$. However, while we had resampled and smoothed the magnetogram data from the various instruments as necessary to account for the differences in spatial resolution (Sect. \ref{preparation}), \cite{yeo14} examined them at their original spatial resolution, which obviously leads to different conversion factors.

\subsection{Application of the results from the current study to earlier studies}
\label{examples}

In this study, we had quantified the uncertainty in measured magnetism due to changes
in instrument response, results that are useful to the interpretation of SPM, MDI, VSM,
and HMI magnetograms. As examples, let us discuss the implications of the current findings on the investigations by \cite{meunier18} and \cite{yeo14}.

\subsubsection{\cite{meunier18}}
\label{examplesm18}

Taking MDI magnetograms from the period of 1996 to 2010, \cite{meunier18} calculated the mean $\ablos/\mu$ over the quietest region near disc centre, $\inmf$, which they interpreted as a measure of the internetwork magnetic field. From 1996 to 1999, $\inmf$ increases steadily from about 6 G to 7.5 G, from 1999 to 2001 it is roughly constant, from 2001 to 2004 it decreases steadily from 7.5 G to 6.5 G and thereafter, it remains roughly constant again (see Fig. 2D in their paper). The observation that $\inmf$ is lower during the 1996 and 2008 solar cycle minima and higher during the 2001 maximum suggests that the internetwork magnetic field varies in phase with the solar cycle, in contradiction to various studies which concluded otherwise \citep{lites11,buehler13,rempel14,lites14}.

The current study revealed that over the period of 1996 to 1999, the response of MDI relative to SPM had risen such that, as compared to SPM, the MDI magnetogram signal appear to be biased upwards by $24\pm2\%$, discussed in Sect. \ref{analysisstability}. As noted then, this is likely the sum manifest of response changes in both instruments and not either instrument alone. The SoHO vacation and the MDI refocus, which occurred during this period, is accompanied by a bias of $15\pm2\%$, suggesting that at least this much of the total bias of $24\pm2\%$ might be due to MDI alone. In other words, the response of MDI changed in this period such that its magnetogram signal is biased upwards by at least 15\% and possibly by as much as 24\%. Comparing this to the 25\% increase in $\inmf$ over the same period reported by \cite{meunier18}, we conjecture that what the author noted is at least in part, if not completely, an instrumental artefact.

\subsubsection{\cite{yeo14}}
\label{examplesyea14}

\cite{yeo14} reconstructed the variation in total and spectral solar irradiance, TSI and SSI, since 1974 using a model based on full-disc magnetograms. They made use of full-disc magnetograms from the KPVT 512-channel magnetograph \citep[512CM,][]{livingston76}, SPM, MDI(2) (i.e. observations from after the SoHO vacation and the refocus of this instrument), and HMI. In order for the model output based on the observations from the various instruments to be mutually consistent, \cite{yeo14} rescaled the MDI(2) magnetograms to the HMI magnetogram signal scale with the conversion factor determined as a function of $\blos$ and $\mu$, similar to the case A analysis we had presented in Sect. \ref{analysisrelationship}. The SPM magnetograms were similarly cross-calibrated to the rescaled MDI(2) magnetograms and the 512CM magnetograms to the rescaled SPM magnetograms.

The solar irradiance reconstruction presented by \cite{yeo14} is based on SPM magnetograms for the period of 1992 to 2003 and MDI(2) magnetograms for the period of 1999 to 2010. The reconstruction indicates that between the 1996 and 2008 solar cycle minima, TSI declined by $0.24\ {\rm Wm^{-2}}$. Considering the fact that the variation in TSI over the solar cycle is around $1\ {\rm Wm^{-2}}$, this is significant. Making use of the results obtained in the current study, let us examine the likelihood that the minimum-to-minimum decline indicated by the reconstruction might be an artefact of systematic differences between the 1996 SPM magnetograms and the 2008 MDI(2) magnetograms.

The \cite{yeo14} model describes the contribution to solar irradiance variability by faculae, identified in the magnetograms, and by sunspots, identified in concurrent continuum images. In the \cite{yeo14} model, to an approximation, a 10\% bias in $\ablos$ is accompanied by a bias of $0.08\ {\rm Wm^{-2}}$ in the TSI reconstruction. For the minimum-to-minimum decline of $0.24\ {\rm Wm^{-2}}$ in the TSI reconstruction to have come from a bias in the 1996 SPM magnetograms relative to the 2008 MDI(2) magnetograms, this bias has to render 1996 SPM magnetogram signals 30\% stronger.

Let us examine if the cross-calibration between SPM and MDI(2) and the time variation in instrument response might have produced such a bias. As noted in the beginning of this section, the \cite{yeo14} cross-calibration of SPM and MDI(2) magnetograms is broadly similar to the case A analysis presented in Sect. \ref{analysisrelationship}, where we determined the uncertainty in $\blos$ due to the cross-calibration, $\rmsblosa$, to be about 5\% (Table \ref{agreement}). This being the case, we expect \cite{yeo14} to have reconciled the magnetogram signal registered by these two instruments to within a similar margin. Looking at Fig. \ref{stability}, at the instrument comparisons that included either SPM or MDI(2), the quantity $\rdblos$ (dashed lines), which traces the relative bias in $\blos$ due to the time variation in instrument response, is at most times below 5\%. For the SPM-MDI(1) comparison, drawn in cyan, there is a steady drift in $\rdblos$ of about 10\% over the period of 1996 to 1998. For the sake of argument, let us suppose that MDI(1) had been stable over this period such that this steady drift is due to SPM alone. This would mean that relative to the 1998 SPM magnetograms, the 1996 SPM magnetograms are biased such that magnetogram signals are 10\% stronger. The cross-calibration and the time variation in instrument response could have contributed to the minimum-to-minimum decline in the TSI reconstruction, but none of these factors appear strong enough (being well below 30\%) to fully account for this drop. We emphasis that this is not a validation of the \cite{yeo14} finding. All this discussion shows is that the minimum-to-minimum drop is unlikely, at least not predominantly, from a bias in the 1996 SPM magnetograms relative to the 2008 MDI(2) magnetograms.

\section{Summary}
\label{summary}

The utility of the full solar disc magnetograms recorded over the past half-century as a long-term record of photospheric magnetism depends on an understanding of how stable these observations are with time and the systematic differences between the various instruments. While studies comparing the observations from the various solar telescopes have found evidence of fluctuations in instrument response with time, magnetogram signal level, and disc position, the effect on measured magnetism has not been examined in detail (Sect. \ref{introduction}). The aim of this study is to address this issue. Comparing full-disc magnetograms from the KPVT/SPM \citep{jones92}, SoHO/MDI \citep{scherrer95}, SOLIS/VSM \citep{keller03}, and SDO/HMI \citep{schou12}, we quantified the uncertainty in the longitudinal magnetogram signal, $\blos$ and disc-integrated photospheric magnetic flux, $\ftot$ due to the variation in instrument response with time, magnetogram signal and distance from disc centre.

For each instrument-to-instrument comparison, we selected magnetograms the two instruments recorded close in time to one another and from over the entire period both of them operated (Sect. \ref{selection}). The data were processed in such a way as to factor out the differences in observation time, spatial resolution, and magnetogram noise level (Sect. \ref{preparation} and \ref{magnetogramnoise}). We compared the surface coverage by magnetic features in the two sets of magnetograms and found the bias we have to introduce to the magnetograms from one of the two instruments in order to offset the difference. This allowed us to trace the time variation in instrument response in terms of its effect on measured $\blos$ and $\ftot$ (Sect. \ref{analysisstability}). This approach circumvents certain limitations of earlier attempts to examine such instrument changes (Sect. \ref{analysis}). We derived the conversion factor, the ratio of the magnetogram signal registered by the two instruments, as a function of $\blos$ and the cosine of the heliocentric angle, $\mu$. Taking this result, we rescaled the magnetograms from one instrument to the magnetogram signal scale of the other. We repeated this analysis with the global conversion factor, in other words, the overall ratio derived ignoring the $\blos$ and $\mu$-dependence. By comparing the results from the two cross-calibrations, we estimated the uncertainty in $\blos$ and $\ftot$ due to the variation in instrument response with magnetogram signal level and distance from disc centre (Sect. \ref{analysisrelationship}). Earlier studies have examined the conversion factor between the same instruments we looked at here, reviewed in Sect. \ref{cfcomparison}, and the study by \cite{riley14} even made use of this quantity to trace the time variation in instrument response. This study extends these earlier works by examining the effect of instrument response changes on measured magnetism in detail.

Between 1998 and 1999, the SoHO spacecraft suffered a series of extended outages, often referred to as the SoHO vacation, at the end of which the MDI instrument was refocused. From 1996, when MDI started operation, right up to the SoHO vacation, the response of MDI, relative to SPM, drifted steadily (Figs. \ref{cover}a and \ref{stability}). The SoHO vacation and MDI refocus is accompanied by a further offset in the relative response (Fig. \ref{exceptional}a). The result is that as compared to SPM, the MDI magnetogram signal appear to be biased upwards by 24±2\% over the period of 1996 to 1999, with 15±2\% coming from the SoHO vacation and MDI refocus alone. By this estimation, MDI magnetograms from before and after the SoHO vacation and MDI refocus can be reconciled to one another by scaling the former by a factor of $1.15\pm0.02$. After the SoHO vacation and MDI refocus, MDI and SPM appear to be relatively stable with respect to one another (Figs. \ref{cover}b and \ref{stability}).

The VSM-to-MDI comparison indicted that the response of VSM might have changed with the relocation of the SOLIS instrument suite in 2004, the modulator degrading from 2003 to 2006 before its replacement, and the CCD camera upgrade in 2009 (Figs. \ref{stability} and \ref{exceptional}b). This is such that as compared to MDI, the VSM magnetogram signal appears to be biased by about 10\% to 20\% with each of these events. From this analysis, we estimated that VSM magnetograms from before the CCD camera upgrade can be cross-calibrated to the post-upgrade observations by scaling them by a factor of $1.18\pm0.02$. When VSM switched from filling its interior with helium to using nitrogen instead in 2014, the spatial resolution degraded slightly. The VSM-to-HMI comparison indicated that the change in the response of VSM relative to HMI this produced appears to be similar in magnitude to the fluctuations that occurred at other times in the regular operation of both instruments (Fig. \ref{exceptional}c).

Apart from what we have just noted about MDI and VSM, the various instruments appear to be relatively stable with respect to one another (Fig. \ref{stability}). The uncertainty in $\blos$ and $\ftot$ due to the time variation in instrument response is less than 2\% (Table \ref{uncertainty}). From the cross-calibration analysis, we estimated the uncertainty in $\blos$ due to the variation in instrument response with magnetogram signal level and distance from disc centre to be about 8$\%$ to 14$\%$ and in $\ftot$ to be 7$\%$ to 23 $\%$ (Table \ref{agreement}). (We did not examine in detail the variation in instrument response with polar angle, i.e. the non-radial component of the variation with disc position, but we did find evidence that it is likely minute in comparison, see Fig. \ref{relationship}.) Taking the variation in instrument response with magnetogram signal and distance from disc centre into account in the cross-calibration brought the various instruments to with 2\% to 5\% of one another in terms of $\blos$ and 5\% to 9\% in terms of $\ftot$. This residual and the uncertainty due to the time variation in instrument response are smaller than the uncertainty due to the variation in instrument response with magnetogram signal level and distance from disc centre, suggesting that the latter dominates the uncertainty in measured magnetism and the systematic differences between the observations from the various instruments.

The results obtained here are useful for the interpretation of SPM, MDI, VSM, and HMI magnetograms, whether a reassessment of earlier results obtained with such data or to inform future studies. As examples, we applied our findings to certain results from \cite{meunier18} and \cite{yeo14}, detailed in Sect. \ref{examples}. \cite{meunier18} reported a rise in the strength of the internetwork magnetic field, as apparent in MDI observations, over the period of 1996 to 1999. We argued it cannot be excluded that this trend is, at least in part, an artefact of the changes in the response of MDI over this period. The model reconstruction of total solar irradiance by \cite{yeo14} indicates a decline in this quantity between the 1996 and 2008 solar cycle minima. The 1992 to 2010 segment of the reconstruction is based on cross-calibrated SPM and MDI magnetograms. From the results of the current study, we argued that the minimum-to-minimum drop reported by \cite{yeo14} is unlikely to be, at least not predominantly, an artefact of the cross-calibration or the time variation in the response of either instrument. %The model of the disc-integrated magnetic flux in faculae and network, $\ftotfn$ by \cite{yeo20} reproduces about 90\% of the variability in measured $\ftotfn$, which is determined from full-disc magnetograms. We showed that the uncertainty in magnetogram observations due to the various sources of error we have examined in this study accounts for a significant proportion of the residual between measured and modelled $\ftotfn$.

\begin{acknowledgements}
The authors are grateful to Jack Harvey and Andrew Marble for the useful discussions. This work made use of magnetogram observations from KPVT/SPM, SoHO/MDI, SOLIS/VSM, and SDO/HMI. This work was supported by the German Federal Ministry of Education and Research (Project No. 01LG1909C) and the European Research Council (ERC) under the European Union’s Horizon 2020 research and innovation program (Grant Agreement No. 695075).

\end{acknowledgements}

\bibliographystyle{aa}
\bibliography{references}

\end{document}